\documentclass[twocolumn]{emulateapj}
\usepackage{amsfonts}
\usepackage{amsmath}
\usepackage{amssymb}
\usepackage{graphicx}
\usepackage{epsfig,lscape}
\usepackage{natbib}

\newcommand\mum{$\mu$m}
\newcommand\hi{H$\,${\small I}}
\newcommand\sfr{$\Sigma_{\rm SFR}$}
\newcommand\shi{$\Sigma_{\rm HI}$}
\newcommand\sighi{$\sigma_{\rm HI}$}
\newcommand\hii{H$\,${\small II}}
\newcommand\halpha{H$\alpha$}
\newcommand\kms{$\rm km\;s^{-1}$}
\newcommand\dg{$^{\circ}$}
\newcommand\msun{M$_\odot$}
\def\avg#1{\langle #1 \rangle}

 \slugcomment{To appear in AJ.}

\shorttitle{What is Driving the {\sc H\,i} Velocity Dispersion?}
\shortauthors{D.Tamburro, et al.}

\begin{document}

\title{What is Driving the H{\sc i} Velocity Dispersion?}

\author{D.~Tamburro\altaffilmark{1}, H.-W.~Rix\altaffilmark{1},
  A.K.~Leroy\altaffilmark{1}, M.-M.~Mac~Low\altaffilmark{2,1},
  F.~Walter\altaffilmark{1}, R.C.~Kennicutt\altaffilmark{3},
  E.~Brinks\altaffilmark{4}, and W.J.G.~de~Blok\altaffilmark{5}}

\email{tamburro@mpia.de, rix@mpia.de, leroy@mpia.de, mordecai@amnh.org,
  walter@mpia.de,  robk@ast.cam.ac.uk, e.brinks@herts.ac.uk,
  edeblok@circinus.ast.uct.ac.za}

\altaffiltext{1}{Max-Planck-Institut f\"ur Astronomie, K\"onigstuhl 17,
 D-69117 Heidelberg, Germany}

\altaffiltext{2}{Department of Astrophysics, American Museum of Natural
  History, 79th Street and Central Park West, New York, NY 10024-5192, USA}

\altaffiltext{3}{Institute of Astronomy, University of Cambridge,
   Madingley Road, Cambridge CB3 0HA, United Kingdom}

\altaffiltext{4}{Centre for Astrophysics Research, University of
  Hertfordshire, College Lane, Hatfield AL10 9AB, United Kingdom}

\altaffiltext{5}{Department of Astronomy, University of Cape
   Town, Private Bag X3, Rondebosch 7701, South Africa}

\begin{abstract}  
  We explore what dominant physical mechanism sets the kinetic energy
  contained in neutral, atomic (\hi) gas.  Supernova (SN) explosions and
  magneto-rotational instability (MRI) have both been proposed to drive
  turbulence in gas disks and we compare the \hi\ line widths predicted from
  turbulence driven by these mechanisms to direct observations in 11 disk
  galaxies.  We use high-quality maps of the \hi\ mass surface density and
  line width, obtained by the THINGS survey. We show that all sample galaxies
  exhibit a systematic radial decline in the \hi\ line width, which appears to
  be a generic property of \hi\ disks and also implies a radial decline in
  kinetic energy density of \hi.  At a galactocentric radius of $r_{25}$ --
  often comparable to the extent of significant star-formation -- there is a
  characteristic value of the HI velocity dispersion of $10\pm 2$~\kms.
  Inside this radius, galaxies show \hi\ line widths well above the thermal
  value (corresponding to $\sim 8$~\kms) expected from a warm \hi\ component,
  implying that turbulence drivers must be responsible for maintaining this
  line width.  Therefore, we compare maps of \hi\ kinetic energy to maps of
  the star formation rate (SFR) -- a proxy for the SN rate -- and to
  predictions for energy generated by MRI.  We find a positive correlation
  between kinetic energy of \hi\ and SFR; this correlation also holds at fixed
  \shi, as expected if SNe were driving turbulence.  For a given turbulence
  dissipation timescale we can estimate the energy input required to maintain
  the observed kinetic energy. The SN rate implied by the observed recent SFR
  is sufficient to maintain the observed velocity dispersion, if the SN
  feedback efficiency is at least $\epsilon_{\rm SN} \simeq
  0.1\times(10^7\;{\rm yr}/\tau_D)$, assuming $\tau_D\simeq10^7$~yr for the
  turbulence dissipation timescale. Beyond $r_{25}$, this efficiency would
  have to increase to unrealistic values, $\epsilon \gtrsim 1$, suggesting
  that mechanical energy input from young stellar populations does not supply
  most kinetic energy in outer disks. On the other hand, both thermal
  broadening and turbulence driven by MRI can plausibly produce the velocity
  dispersions and kinetic energies that we observe in this regime ($\gtrsim
  r_{25}$).

\end{abstract}


\maketitle

\section{INTRODUCTION}

The extended \hi\ disks of spiral galaxies exhibit characteristic 21 cm line
widths of 5--15\ \kms\ on small scales.  Line widths are observed to be
$\sim6$--10~\kms\ even beyond $r_{25}$, where the star formation rate (SFR)
and the associated feedback are presumably inefficient
\citep*{vanderkruit1982, Dickey1990, Kamphuis1993, Rownd1994, vanzee1999}.  In
some observed cases, the \hi\ disks exhibit an overall outward decrease of the
velocity dispersion to 6--8\ \kms\ \citep{Boulanger1992,Petric2007}.

The observed line widths are attributable in part to thermal broadening and in
part to turbulence.  At pressures typical of disk galaxies, interstellar
atomic gas has two temperatures at which it can remain in stable thermal
equilibrium \citep{Wolfire1995}. The gas is observed to be distributed with
roughly 60\% of the mass evenly distributed between the equilibrium
temperatures \citep{Kulkarni1987, DickeyLockman1990, Dickey1993} and 40\%
transiting between them \citep{Heiles2001}.  The cold ($\sim100$~K) neutral
gas emits lines with a characteristic line width of $\sim1$~\kms, while the
warm ($\sim8000$~K) neutral gas has a line width of $\sim8$~\kms.  Spectral
lines observed to be broader than $\sim8$~\kms\ can be attributed to
turbulence -- stochastic fluctuations of velocity and pressure on scales much
larger than the mean free path of the gas.  Heating by a galactic interior or
extragalactic background of ultraviolet (UV) radiation \citep{Schaye2004} may
be significant in the outermost regions of \hi\ disks, where velocity
dispersions approach thermal values for the warm phase.
 
There are several energetic sources that can induce turbulent motions in the
interstellar medium (ISM), among which feedback from recently formed stellar
populations and magnetorotational instabilities
\citep[MRI;][]{Balbus1991,Balbus1998,Sellwood1999} are some of the most
important.  The stirred turbulence drives motions at smaller and smaller
scales until viscosity dissipates the motion into thermal energy.  This
hierarchical description was formulated by \citet{Kolmogorov1941} for an
incompressible turbulent fluid, though it applies well to the small scale
dynamics of the gas in galaxy disks \citep*{Elmegreen2003}.  Ultimately, the
ISM radiates away the thermal energy, e.g, by line cooling.  Since the typical
radiative cooling time \citep[$\sim10^3$~years;][]{Wolfire2003} is shorter
than the time scale for turbulent energy to decay
\citep[$\sim10$~Myr;][]{maclow1999}, the turbulent energy must be continuously
replenished on the decay time scale.  The direct identification of the energy
sources that sustain the turbulence at this rate has drawn considerable
interest for well over three decades, but has not yet been concluded
\citep[e.g.][]{maclow2004, elmegreen2004}.

Numerical simulations have demonstrated that both hydrodynamical and
magnetohydrodynamic turbulence decay over a relatively rapid time scale,
comparable to the local dynamical time, i.e. $\sim 10$~Myr \citep*{maclow1998,
  Stone1998, maclow1999, Padoan1999, Avila2001, Ostriker2001}, thus requiring
a driving mechanism that continuously provides energy to the medium to
maintain a steady state \citep{Gammie1996, maclow1999}.

Which mechanisms drive ISM turbulence remains a matter of debate.  As first
envisioned by \citet{Spitzer1978}, star formation itself injects energy back
to the ISM and drives turbulence through stellar winds of O/B and Wolf-Rayet
stars, UV ionizing radiation, and supernova (SN) explosions
\citep*[e.g.][]{Chiang1985, Rosen1993, Kim2001, Dib2006, Joung2006,
  Avillez2007}. The energy input from SNe likely dominates stellar feedback
\citep{Kornreich2000, maclow2004}.  The overall galaxy kinematics can also
generate some turbulence through spiral wave shocks \citep{roberts69},
rotational shear \citep{Schaye2004}, swing amplified shear instabilities
\citep*{Huber2001,Wada2002}, gravitational potential energy of the arms
\citep{Elmegreen2003}, and non-linear development of gravitational instability
\citep*{li05,li06}.  Magnetized disk instabilities including the
magnetorotational instability \citep*{Balbus1991, Balbus1998,
  Sellwood1999,Dziourkevitch2004, Piontek2005}, and the magneto-Jeans
instability \citep{Wada2004, Kim2006, Kimetal2006}.  Finally, there could also
be a contribution from thermal instability \citep{
  Koyama2002,Kritsuk2002,Dib2005,Hennebelle2007}.

In this paper we attempt a direct observational identification of the energy
sources that drive the turbulence. Our results suggest that the energy input
from young stars determines the \hi\ line width within the star forming region
of galactic disks, while at large radii, where star formation is absent or
feeble, the line width is either sustained by the MRI \citep{Sellwood1999}, or
by thermal broadening from UV heating.  We examine whether SN feedback, gas
temperature, and MRI can maintain effectively the observed ISM line width,
assuming that the energy loss occurs on the self-dissipation timescales
predicted for the turbulence \citep[][]{maclow1999}.  \citet{maclow2004}\ 
argue that SNe, among other mechanisms of energy injection, are the most
effective in actively star forming regions, while in the outermost regions of
galaxy disks, where the SFR is low or zero, the MRI becomes important in
sustaining both the magnetic field, even in low density regions, and the gas
velocity dispersion \citep{Sellwood1999}.

We use multi-wavelength data in the radio, infrared (IR), and far UV
(\S~\ref{sec:data}), to compare detailed maps of the ISM kinetic energy
density implied by 21~cm line widths to maps of the SFR density for a sample
of nearby spiral and dwarf galaxies. In this comparison, we take the SFR
density as a proxy for the energy input rate to the ISM.  Relying on the
measurement of the velocity dispersion as a tracer of turbulence in the
neutral gas, we look at both the correlation between SFR and turbulent motions
in the \hi\ on a pixel-by-pixel basis and at their mean values as function of
galactocentric distance (\S~\ref{sec:analysis}).  We find that the azimuthal
average of the \hi\ velocity dispersion decreases radially much more slowly
than the SFR (\S~\ref{sec:results}); specifically, the radial profile of \hi\ 
velocity dispersion approaches the \hi\ thermal value for the warm neutral
medium where the radial profile of SFR truncates.  Therefore, we investigate
(1) pixel-by-pixel how the kinetic energy of the gas turbulence correlates
with the SFR, especially in terms of SN rate, and (2) what role thermal
broadening and MRI play in regions of low SFR and considerable velocity
dispersion (\S~\ref{sec:discussion}). In particular, we compare the current
turbulent kinetic energy of the gas turbulence to the energy input from both
SNe and MRI (and the efficiencies of this input), on the basis that the
turbulence dissipation timescales, provided by extensive theoretical modeling,
approximate the energy loss timescales.

\section{DATA}\label{sec:data}

The main observational data for our analysis are maps of the ISM kinetic
energy surface density derived from high resolution \hi\ data.  For a
sub-sample we include carbon monoxide (CO) emission maps to explore the
contribution of the molecular gas to the kinetic energy budget. These maps are
then compared, pixel by pixel, to analogous maps of the SFR surface density,
derived from UV and thermal IR maps for the same objects.

We use maps produced by The \hi\ Nearby Galaxy Survey
\citep[THINGS;][]{walter2008}, a survey of the 21 cm emission line carried out
with the NRAO\footnote{The National Radio Astronomy Observatory is a facility
  of the National Science Foundation operated under cooperative agreement by
  Associated Universities, Inc.}\ Very Large Array, to obtain the column
density \shi\ and velocity dispersion \sighi\ of neutral hydrogen.  The THINGS
data have an angular resolution of $\sim7''$, and a velocity resolution of
either 2.6 or 5.2~\kms\ (see Table~\ref{tab:objs}), and a $3\,\sigma$
sensitivity corresponding to a column density of $N({\rm HI})\ge 3.2
\times10^{20}$~cm$^{-2}$. To avoid projection effects in the velocity
dispersion maps, we consider only those galaxies that are more face-on than
$\sim 50$\dg\ \citep[see, e.g.,][]{Leroy2008}, leaving 11 disk galaxies in
total, including three dwarf galaxies (Holmberg~II, IC~2574, and NGC~4214),
listed in Table~\ref{tab:objs}.

We complement the analysis of the \hi\ gas with the mass surface density and
velocity dispersion of molecular hydrogen (H$_2$) gas for about half of our
galaxies: NGC~628, NGC~3351, NGC~3184, NGC~4214, NGC~4736, and NGC~6946.  The
H$_2$ maps are derived from CO $J=2\rightarrow1$ emission maps, obtained at
the IRAM 30-m telescope by \citet{Leroy2008b}\ using the HERA focal plane
array. These maps have resolution of $\sim11''$ and 2.6~\kms, and sensitivity
of $\Sigma_{\rm H2}\ge 4$~\msun~pc$^{-2}$. To derive $\Sigma_{\rm H2}$ we
assume a conversion factor $X_{\rm CO}=2\times10^{20}$~cm$^{-2}$~K~\kms\ and a
line ratio $N(J=2\rightarrow1)/N(J=1\rightarrow0)=0.8$.

The SFR density maps were derived by \citet{Leroy2008} from 24~\mum\ and
far-UV (FUV) maps, combining dust-enshrouded star formation activity and
unobscured star formation, respectively.  The MIPS 24~\mum\ emission maps
\citep{Rieke2004}\ are taken from the Spitzer Infrared Nearby Galaxies Survey
\citep[SINGS;][]{kennicutt03}, and the FUV emission maps are taken from the
Galaxy Evolution Explorer Nearby Galaxy Survey
\citep[GALEX-NGS;][]{gildepaz2007}, for a set of disk galaxies in common with
the THINGS sample.  The MIPS instrument has a resolution of $\simeq6''$ at
24~\mum, and a wide areal coverage, which typically covers $\sim2\,r_{25}$.
The GALEX FUV images have a resolution (FWHM) of $5.6''$ within the
$\lambda=1350-1750$~\AA\ band and have a field of view of $\sim1.25$\dg.

\section{DATA ANALYSIS}\label{sec:analysis}

As a first step in comparing the kinetic energy of the gas to possible drivers
of turbulence, we convert the previously described maps (\S~\ref{sec:data}) of
\hi\ and H$_2$ gas column density and kinematics, and the 24~\mum~+~FUV
surface brightness into physical quantities. Specifically, we obtain mass
surface density, \shi, $\Sigma_{\rm H2}$, velocity dispersion, \sighi,
$\sigma_{\rm H2}$, and SFR surface density, \sfr.  Afterwards, we proceed with
a comparison between the turbulence kinetic energy, obtained from the gas mass
surface density and velocity dispersion, and the SFR on a pixel-by-pixel
basis.

\subsection{Mass Surface Density and Velocity Dispersion}\label{sec:moments}

We determine the \hi\ mass surface density, \shi, and the velocity dispersion,
\sighi, by taking the moments of fully reduced data cubes for our 11 sample
galaxies \citep[the prior reduction steps of are described in][]{walter2008}.
We obtain the pixel-by-pixel maps of $\Sigma_{\rm HI}(x,y)$, $\bar{v}(x,y)$,
and $\sigma(x,y)$, by calculating integrals over the data cubes for all the
positions on the sky $S_i(x,y,v_i)$.  All the sums are calculated over those
channels with signal identified by the THINGS masking \citep{walter2008}; we
only calculated moments for those ($x,y$) line-of-sight positions with signal
in at least 3 channels.

The zero-th moment, $\Sigma_{\rm HI}$, is calculated by integrating the \hi\ 
data cube along the velocity dimension for a total number $N$ of velocity
channels using:
\begin{equation}  \label{eq:m_zero}
  \Sigma_{\rm HI} =  \sum_{i=1}^N S_i,
\end{equation}
where $S_i$ denotes the signal within the $i$-th velocity channel. Afterwards,
\shi\ is corrected for inclination; the adopted inclinations are listed in
Table~\ref{tab:objs}.  For each sample galaxy, the spectral dimension of the
\hi\ data cubes is sufficiently broad to encompass all relevant velocity
values.

The first moment map, the intensity-weighted mean velocity along the line of
sight, is given by:
 \begin{equation}\label{eq:m_one}
\bar{v} =\frac{1 }{\Sigma_{\rm HI}}\;\sum_{i=1}^N v_i\,S_i,
\end{equation}
where $v_i$ is the velocity value of the $i$-th channel. We only use $\bar{v}$
in this context to calculate the central 2nd-moment.

The variance, corresponding to the square of the line-of-sight velocity
dispersion, is given by the intensity-weighted mean deviation:
 \begin{equation}\label{eq:mom_two}
\sigma^2
=\frac{1 }{\Sigma_{\rm HI}}\;\sum_{i=1}^N  (v_i-\bar{v})^2\,S_i.
\end{equation}

Throughout our analysis we restrict ourselves to the moment definition of
Eq.~\ref{eq:mom_two}\ to define the variance.  We presume that $\sigma$
(Eq.~\ref{eq:mom_two}) is the most sensible tracer of turbulent or more
precisely, disordered motions, since it is applicable to any line profile.  We
address the discussion of this assumption in \S~\ref{sec:realities}.

We complement the \hi\ maps with H$_2$ column density and kinematics data
obtained from CO emission (\S~\ref{sec:data}). Following the same procedure
previously adopted to calculate the relevant \hi\ maps, we obtain the mass
surface density, $\Sigma_{\rm H2}$, and the velocity dispersion, $\sigma_{\rm
  H2}$, for the H$_2$ gas.  As described in the following sections, while
H$_2$ is an important contributor to the total mass surface density, its
contribution to the total turbulent kinetic energy of gas turns out to be
rather small in our sample.

\subsection{Realities of Measuring the Velocity 
  Dispersion in THINGS}\label{sec:realities}

In the case of perfect signal, the second moment is exactly the kinetic
energy of \hi\ gas. Even at limited signal-to-noise ratio ($\rm S/N$) but with
normally-distributed noise, we expect the measured variance to scatter about
the true value in a well-behaved way, so that by averaging measurements over
enough area, we can approximate the true value.

As an alternative approach, assuming and then fitting a fixed line profile
shape (e.g., a Gaussian) fails in two ways. First, towards the centers of
galaxies, line profiles tend to be broad and often contain multiple emission
components. Second, in the outskirts of galaxies, where individual lines of
sight tend to have relatively low $\rm S/N$, a simple Gaussian fit tends to
underestimate the line width by latching onto only the peak of the emission.
In the low $\rm S/N$ regime, the line profiles are approximately Gaussian.
For a few sample galaxies (NGC~3184, NGC~5194, NGC~6946, and NGC~7793) we find
that the Gaussian $\sigma$ value is systematically lower than the second
moment. For these galaxies, however, the difference between Gaussian $\sigma$
and 2nd-moment is lower than $\sim 2-3$~\kms\ in average. Since we do not know
a priori the \hi\ gas line-of-sight velocity distribution, and the profiles
are not necessarily Gaussian, the best tracer of kinetic energy, as a model
independent estimate, is the variance.

Therefore, we avoid fitting and attempt to minimize systematics by measuring
the second moment only on regions that have been identified to contain signal.
The production of these ``blanked cubes'' is described in \citet{walter2008},
and is briefly summarized here. First, the cubes were smoothed to 30'' angular
resolution. Then, all areas without $\rm S/N> 2$ in 2 consecutive velocity
channels are blanked (set to 0). This does an excellent job of identifying all
areas with significant emission but still left some false positives. These
were removed by a by-eye inspection that focused on agreement with the overall
velocity structure and morphology of the galaxy. We test the systematics
induced by such blanking on a face-on galaxy by measuring $\sigma$ on the
blanked cube and then expanding the blanking mask and remeasuring $\sigma$. We
find that small expansions of the mask on both directions along the velocity
dimension, e.g. by $\pm 15$~km~s$^{-1}$, have almost no impact on the measured
dispersion. Larger expansion begin to show systematic effects that we want to
avoid (particularly at low $\rm S/N$).

These blanked cubes minimize the effect of artifacts in the THINGS cubes and
yield a more robust estimate of kinetic energy than fitting an assumed line
profile. Therefore, the profiles of $\sigma$ and $E_k$ represent the best
possible estimates using the available data. However, we emphasize that a
rigorous attempt to match other observations or simulations to our
measurements should bear in mind (and ideally duplicate or simulate) this
blanking procedure.

\subsection{Turbulent Kinetic Energy}

With the moment definition of \shi\ and \sighi, the \hi\ kinetic energy per
unit area is given by $E_k=3/2\;\Sigma_{\rm HI}\;\sigma_{\rm HI}^2$, where the
factor 3/2 takes into account all three velocity components assuming that the
velocity dispersion is isotropic.  For the galaxies with molecular gas maps,
we define the total kinetic energy as $E_k=3/2\times(\Sigma_{\rm
  HI}\;\sigma_{\rm HI}^2+\Sigma_{\rm H2}\;\sigma_{\rm H2}^2)$.  In
\S~\ref{sec:results}\ we describe the relationship between the \hi\ (or
\hi~+~H$_2$) gas kinetic energy $E_k$ and the \sfr\ maps in a pixel-by-pixel
scatter plot.

\subsection{Star Formation Rate}\label{sec:limitations}

We use maps of SFR surface density, \sfr, derived by \citet{Leroy2008}, from
combining the 24~\mum\ and the FUV emission maps taken from the SINGS and the
GALEX-NGS surveys (\S~\ref{sec:data}).  \citet{Leroy2008}\ have calibrated the
SFR represented by the UV and IR emission by comparing these maps to \halpha,
24~\mum, and Pa$\alpha$ emission from \hii\ regions\ and young compact stellar
clusters \citep{calzetti07}.  The 24~\mum\ band emission is mostly radiation
from hot dust heated by the UV light from young massive stars and therefore
traces dust-enshrouded, ongoing star formation over a time scale $3-10$~Myr
\citep{Calzetti2005,perez2006,tamburro2008}, although part of the emission
proceeds from outside the \hii\ regions, thus tracing older ($>10$~Myr)
stellar populations.  FUV emission is mostly photospheric emission from O and
B stars. It thus complements the 24~\mum\ emission in regions poor in dust
content, probing therefore low-metallicity and older regions of star formation
over timescales of $\tau\sim10-100$~Myr \citep{Calzetti2005,Salim2007}.

The calibration provides SFR estimates with an uncertainty of 40\% at most at
high SFR, depending on variations in geometry, dust temperature, and age of
stellar populations \citep{Leroy2008}.  At low SFR, a substantial part of the
24~\mum\ emission proceeds from diffuse dust in the ISM, which may not be
associated directly to recent star formation, with the dust being heated by
nearby (young and old) star clusters.  \citet{Leroy2008} argue that below a
fiducial threshold of $\Sigma_{\rm SFR}=10^{-10}\; \rm
M_\odot\;yr^{-1}\;pc^{-2}$ the SFR maps represent upper limits to the true SFR
because of the contribution of this diffuse dust component.

\section{RESULTS}\label{sec:results}

In this section we describe our two main results: (1) the observation that a
radial decline of the \hi\ velocity dispersion is pervasive throughout the
entire sample with little dependence on galaxy type; and (2) that we find
correlation between the kinetic energy of gas and SFR with a slope close to
unity and a similar proportionality constant in all objects.  Both results
depend on the exceptional quality of the data in terms of spatial and velocity
resolution and the wide field of view. We reserve the broader interpretation
of our results for \S~\ref{sec:discussion}.

In the following, we use units of $r_{25}$, which provides a convenient
normalization. \citet{Leroy2008, tamburro2008}\ measured the exponential scale
lengths of near-IR emission from our sample, a good proxy for stellar mass.
$r_{25}$ is typically $4.6\pm0.8$ times the near-IR scale length, and the SFR
maps yield comparable scale lengths to the near-IR. When we normalize by
$r_{25}$ then, it is roughly equivalent to normalizing by the scale length of
the disk.

\subsection{Radial Profiles of $\sigma_{\rm HI}$}\label{sec:radial}

\begin{figure*}
\epsscale{1.0}
\plotone{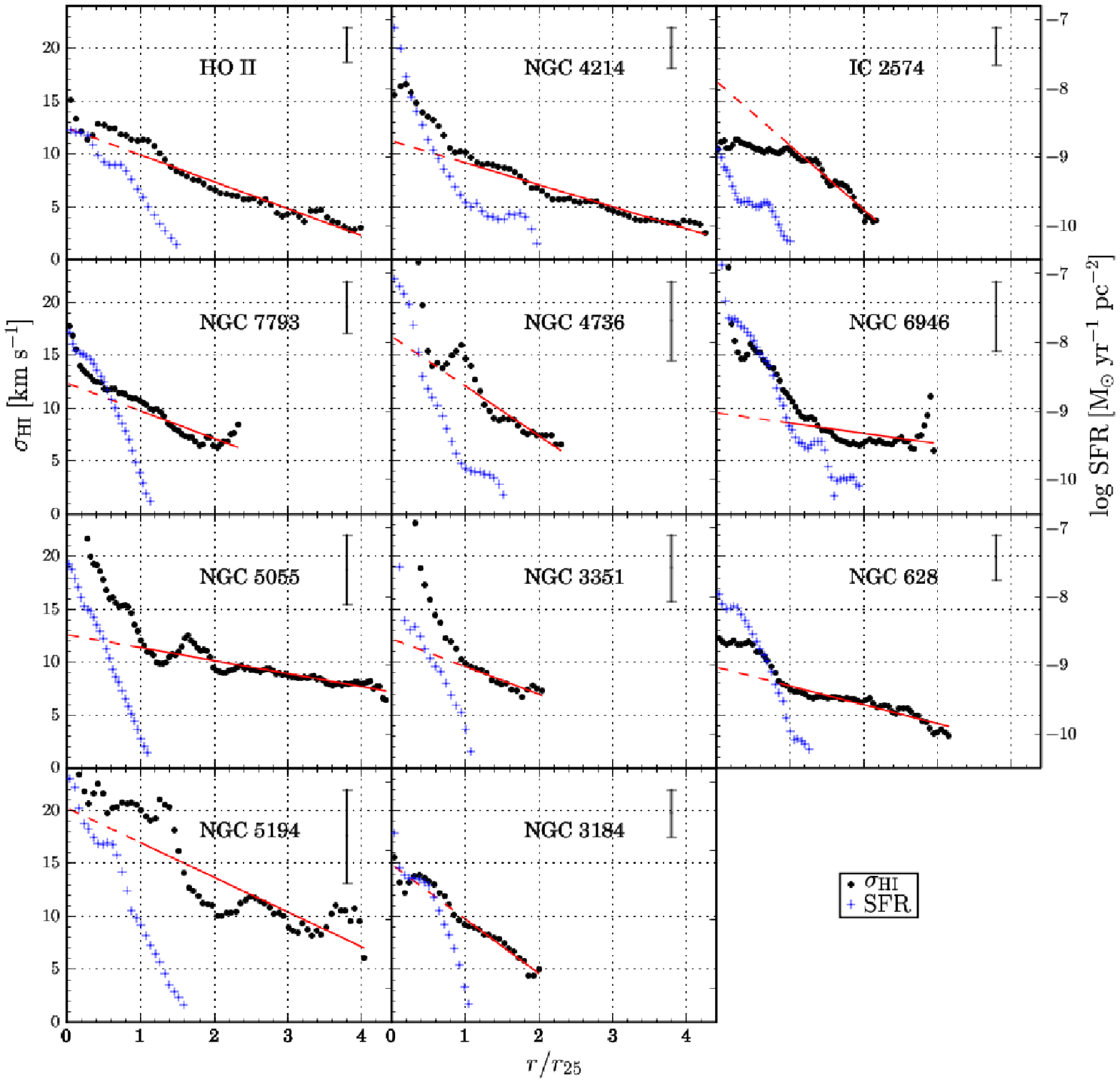}
     \caption{Radial profiles of {\sc Hi} velocity dispersion, \sighi, for the
       galaxies of our sample (\S~\ref{sec:radial}). The filled circles
       represent azimuthal averages of \sighi\ calculated from the second
       moment (Eq.~\ref{eq:mom_two}) as a function of radius $r$ in units of
       $r_{25}$. The error bar on the top right of each panel shows the
       average standard deviation of azimuthal scatter of \sighi.  The blue
       crosses denote the radial profiles of $\log$~\sfr\ with the scale
       indicated on the right side of the plot. Here, only those values above
       the noise threshold ($\log\Sigma_{\rm SFR}/[{\rm
         M_\odot\;yr^{-1}\;pc^{-2}}]\ge-10$) are plotted
       (\S~\ref{sec:limitations}). The labeled galaxies are sorted by
       increasing dynamical mass \citep[$v_{\rm max}$][]{deblok2008,
         tamburro2008}.}\label{fig:m2_vs_r_fit}
\end{figure*}

To start, we determine the azimuthally averaged radial profiles of \hi\ 
velocity dispersion, SFR, and gas kinetic energy, by calculating the average
values of \sighi\ and SFR within annuli of $\sim15''$ width.  The resulting
\sighi$(r)$ and \sfr$(r)$ profiles are shown in Fig.~\ref{fig:m2_vs_r_fit}.
They exhibit a radial decline of \hi\ velocity dispersion as a common
characteristic for all sample galaxies independent of their dynamical mass.
While previous studies have reported this individually for a few disk galaxies
\citep[i.e.,][]{Boulanger1992, Petric2007}, we show for the first time that
this is true for a significant sample of dwarf and normal spiral galaxies.

\begin{figure*}
\epsscale{0.7}
\plotone{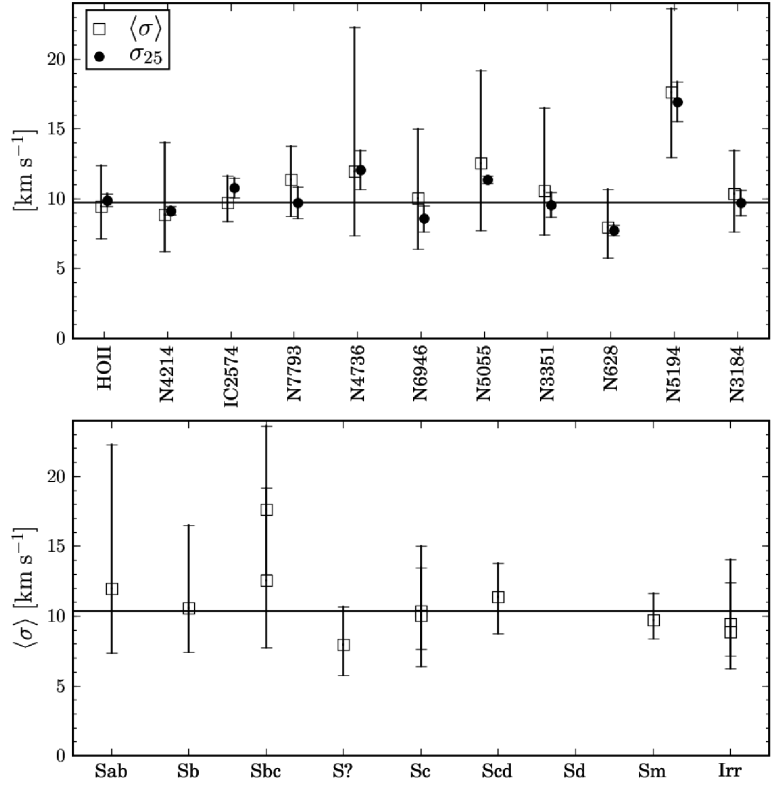}
\caption{Mean values of {\sc Hi} velocity dispersion for our sample galaxies
  (\S~\ref{sec:radial}). Top panel: the filled circles and the corresponding
  error bars represent the value of the {\sc Hi} velocity dispersion at
  $r_{25}$, $\sigma(r_{25})$, from the linear fit in Fig~\ref{fig:m2_vs_r_fit}
  (\S~\ref{sec:radial}); the squares represent the \shi-weighted median
  $\avg{\sigma}$ with the error bars indicating the 68\% confidence interval
  (cf. Table~\ref{tab:objs}). The solid horizontal line represents the median
  value ($10\pm2$~\kms) of all $\sigma(r_{25})$ for the sample galaxies. The
  labeled galaxies are sorted by increasing dynamical mass ($v_{\rm max}$).
  Bottom panel: the squares represent the \shi-weighted median $\avg{\sigma}$
  as in the top panel; the galaxies are sorted by Hubble type (cf.
  Table~\ref{tab:objs}). The solid horizontal line represents the median of
  all $\avg{\sigma}\simeq10\pm2$~\kms.  }\label{fig:s25_avg}
\end{figure*}

To characterize the \sighi\ radial gradients, we fit a linear relation to the
radial profiles of \sighi\ for all $r\ge r_{25}$, a regime where $\sigma(r)$
is approximately linear with $r$ for all the galaxies of the sample. With
$\sigma(r)$ going from $\gtrsim20$~\kms\ down to $\sim5\pm2$~\kms\ near the
outermost observed radius (Fig.~\ref{fig:m2_vs_r_fit}), the velocity
dispersion decreases with radius by $\simeq3$--5~\kms\ per $\Delta r_{25}$;
for comparison, azimuthal variations at a fixed radius are $\simeq 5$~\kms.
In Fig.~\ref{fig:s25_avg}, we display and summarize the intercept value of
\sighi\ at $r_{25}$ and the \hi\ mass weighted median of \sighi\ for all
sample galaxies individually. Our observations resolve the outward radial
\sighi\ decline well, since a typical $r_{25}$ for our sample galaxies
corresponds to $\sim30$--40 times the 7'' \hi\ resolution limit and the
typical radial extent of the \hi\ emission is $2-4\times r_{25}$.

Remarkably, we find that all galaxies have the same \hi\ velocity dispersion
at their respective $r_{25}$: a general value of
$\sigma(r_{25})\simeq10\pm2$~\kms, which displays no apparent trend with the
dynamical mass and morphological type, and is consistent with the \hi\ 
mass-weighted median value $\langle\sigma\rangle$ (see Fig.~\ref{fig:s25_avg}\ 
and Table~\ref{tab:objs}).  The radial \sighi\ gradient has in all cases the
same sign as the much steeper decline of the mean SFR as function of radius as
shown for comparison in logarithmic scale in Fig.~\ref{fig:m2_vs_r_fit}.

Since the sample galaxies are more face-on than 50\dg, the systematic increase
of \sighi\ towards the center cannot be due to increasing beam smearing
effects.  We analyze the combined effect of beam smearing, inclination and
rotation curve on the velocity dispersion by constructing a sample data cube
containing signal in only one velocity channel per each ($x,y$) spatial
position (i.e. intrinsic \sighi~=~0) and characterized by the same inclination
and rotation curve of NGC~5055 -- the most inclined disk of the sample with a
fast rotation speed.  After convolving this sample data cube with a kernel of
7'' FWHM (the resolution of our \hi\ maps), we calculate the velocity
dispersion using second moments (Eq.~\ref{eq:mom_two}) as done throughout our
analysis (\S~\ref{sec:moments}).  The resulting line broadening from beam
smearing for a disk with the rotation curve and orientation of NGC~5055 is
only $\lesssim 5 $~\kms, lower than the observed velocity dispersion in
NGC~5055 by $\sim10$--20~\kms.  The fractional contribution of the velocity
dispersion from beam smearing to the total observed velocity dispersion is
only 20\% in the central region ($r<1/4\; r_{25}$) and at most 10\% at larger
galactocentric radii.

\subsection{The H{\sc i} Kinetic Energy Density as a Function of Radius}\label{sec:radialek}

The \hi\ kinetic energy density, $E_k$, in each pixel exhibits a clear radial
decline, as shown in the full pixel-by-pixel distribution
(Fig.~\ref{fig:Ek_vs_r}), where the black contours and color scale indicate
the density of pixels at each \hi\ kinetic energy and radius, while the red
contours show the sum of atomic and molecular kinetic energy.  The latter
diverges from the \hi\ kinetic energy in the inner parts of some galaxies,
indicating that the cold molecular gas contributes to the kinetic energy
budget. Since every galaxy of the sample includes $\sim2\times10^5$ pixels
with significant signal, we display in Fig.~\ref{fig:Ek_vs_r} the density
contours of the data points.

We include the analysis of the H$_2$ mass surface density and velocity
dispersion derived from the CO emission for a few sample galaxies
(\S~\ref{sec:analysis}), to quantify the contribution of the molecular gas to
the total kinetic energy at high H$_2$-to-\hi\ mass ratio.  The total kinetic
energy $E_k=E_{\rm HI}+E_{\rm H2}$ is plotted in Fig.~\ref{fig:Ek_vs_r}\ (with
red contours) for the galaxies with CO data. For the galaxies NGC~4736,
NGC~5055, and NGC~6946, $E_{\rm H2}$ is comparable to $E_{\rm HI}$ or even
dominant in the central regions of galaxy disks; for these galaxies the H$_2$
mass and spatial extent is considerable. For those galaxies where the \hi\ gas
dominates the total gas mass, i.e., NGC~628, NGC~3184, NGC~3351, and the dwarf
galaxies Holmberg~II, IC~2474, and NGC~4214, characterized by little or no
detected molecular gas, the molecular gas does not contribute much to the
total kinetic energy.

\subsection{Correlation between $E_k$ and \sfr}\label{sec:pixel}

In Fig.~\ref{fig:Ek_vs_sfr} we compare pixel-by-pixel the relation between the
kinetic energy density of the \hi\ gas, $E_k=3/2\,\Sigma_{\rm HI}\,\sigma_{\rm
  HI}^2$, and the SFR surface density -- a proxy for the energy input rate by
SN.  We find that in all galaxies these quantities are well correlated with a
slope close to unity and with no evident dependence on dynamical mass
(Fig.~\ref{fig:Ek_vs_sfr}).  Note that a considerable fraction of all data
points in Fig.~\ref{fig:Ek_vs_sfr}\ lie below the noise estimated for the
\sfr\ maps (\S~\ref{sec:limitations}). These data points lie in the outermost
parts of galaxy disks ($r>2\times r_{25}$), where there is little or no
ongoing star formation (cf. Fig.~\ref{fig:Ek_vs_r}).

We note that the slope of all correlations in Fig.~\ref{fig:Ek_vs_sfr}\ 
flattens at high $E_k$ and \sfr, although we argue in \S~\ref{sec:radialek}
that this is not caused by a higher abundance of molecular gas at higher \sfr.
The total kinetic energy $E_k=E_{\rm HI}+E_{\rm H2}$ is plotted in
Fig.~\ref{fig:Ek_vs_sfr} (with red contours) for the six sample galaxies with
H$_2$ data. Only for NGC~5055 and NGC~6946 is the kinetic energy of the
molecular gas, $E_{\rm H2}$ important; in those cases the $E_{\rm HI}+E_{\rm
  H2}$ vs.\ \sfr\ relation is linear and the slope is close to unity.

\section{DISCUSSION}\label{sec:discussion}

What scenario does our data support for producing the observed line widths?
SN-driven turbulence seems likely to be the dominant factor broadening
line widths within the radius of active star formation, since we find that the
level of predicted SN energy is sufficient to account for the turbulent
kinetic energy implied by the line width as a function of radius.

The radial slopes of SN energy and kinetic energy of the neutral (\hi\ and
H$_2$) gas agree qualitatively so that the kinetic energy of the gas is
proportional to the local star formation rate.  Yet, the fact that the \hi\ 
velocity dispersion approaches its thermal value of roughly 6~\kms\ well
beyond the radius of detectable star formation indicates that either (1) the
line broadening is due to UV heating, with a warm neutral medium temperature
of $\sim 5000$~K resulting in \sighi~$\sim 6$~\kms, or (2) the gas is actually
turbulent and another mechanism such as the MRI is driving the turbulence.  We
now explore whether our new data support this scenario.

In the following, we compare the observed line widths with the most plausible
mechanisms for generating them: SNe (\S~\ref{sec:supernova}), UV heating
(\S~\ref{sec:thermal}), and MRI (\S~\ref{sec:mri}). SNe and MRI both produce
broad line widths by driving turbulence, while UV heating can produce thermal
broadening.  The required energy injection rate depends both on the kinetic or
thermal energy of the gas, which can be derived from the observed line widths
and the turbulence decay or cooling timescales, which must be derived from
models.  More precisely, the gas kinetic energy implied by the linewidth
consists of a combination of turbulence and the thermal energy associated with
the (warm) gas temperature, i.e. $E_k=E_{\rm turb}+E_{\rm therm}$.  If the
thermal broadening is much less effective than the turbulence, then $E_k\simeq
E_{\rm turb}$.

If the gas turbulence is mainly driven by SNe and MRI, then we expect the
dissipation rate of turbulence to equal the sum of the energy input rates of
SNe and MRI,
\begin{equation} 
\dot{E}_k \simeq \epsilon_{\rm SN}\,E_{\rm
  SN} /\tau_{\rm SN}+ \epsilon_{\rm MRI} E_{\rm MRI} / \tau_{\rm MRI},
\end{equation}
where $\epsilon_{\rm SN}$ and $\epsilon_{\rm MRI}$ are the efficiencies, and
$\tau_{\rm SN}$ and $\tau_{\rm MRI}$ are the decay times of turbulence driven
by the two mechanisms.  Different mechanisms can result in different decay
rates because they have different driving scales and magnitudes
\citep{Stone1998,maclow1999}.

\subsection{Supernova Energy}\label{sec:supernova}

Assuming steady state equilibrium between the energy input rate from SNe to
turbulent gas motions and the energy loss rate from dissipation of this
turbulence, then the resulting kinetic energy $E_k = \epsilon_{\rm SN}
\dot{E}_{\rm SN} \tau_{SN}$, where $\dot{E}_{\rm SN}$ is the rate of released
SN energy, which we estimate from the SFR, and the SN feedback efficiency
$\epsilon_{\rm SN}$ is the fraction of SN energy converted to turbulent
motions in the cold gas.  \citet{maclow1999} finds that the dissipation rate
of turbulence depends on the driving scale $\lambda$ and the velocity
dispersion $\sigma$ as
\begin{equation}
\tau_D\simeq9.8\;(\lambda_{100}/\sigma_{10})\;\rm Myr,
\end{equation}
where $\lambda_{100} = \lambda / 100$~pc and $\sigma_{10} = \sigma /
10$~km~s$^{-1}$.  Numerical simulations of SN-driven turbulence yield
$\lambda=100\pm30$~pc \citep{Joung2006, Avillez2007}, and our own analysis
gives an average velocity dispersion $\sigma=10$~\kms (\S~\ref{sec:radial}).
The SN energy input rate, $\dot{E}_{\rm SN}$, can be estimated from the SN
rate implied by our SFR maps.  The SN rate per unit area, $\eta$, depends on
the fraction $f_{*\rightarrow \rm SN}$ of all recently formed stars that
terminate in core-collapse SNe:
\begin{equation}
  \label{eq:sn_rate}
\eta= \frac {\rm SFR}{\avg{m}}  \times  f_{*\rightarrow \rm SN},
\end{equation}
where $\avg{m}$ is the average mass of stars of the population.  We assume
that only those stars in the mass range ($8-120$)\,\msun\ can form
core-collapse SNe. The SFR maps used in our analysis 
  assume an initial mass function (IMF)
$\phi(m)=m^{-\alpha}$, where $\alpha=1.3$ for the mass range
($0.1-0.5$)\,\msun\ and $\alpha=2.3$ for the mass range ($0.5-120$)\,\msun\ 
\citep[cf.][]{Leitherer1999, calzetti07}.  Then, the SN fraction
\begin{equation}
  \label{eq:frac_sn}
  f_{*\rightarrow \rm SN}/\avg{m}=
\frac{\int_{8\,\rm M_\odot}^{120\,\rm M_\odot}
  \phi(m)\, dm} {\int_{0.1\,\rm M_\odot}^{120\,\rm M_\odot}
  \phi(m)\, dm} / \frac{\int_{0.1\,\rm M_\odot}^{120\,\rm M_\odot}
  m \phi(m)\, dm} {\int_{0.1\,\rm M_\odot}^{120\,\rm M_\odot}
  \phi(m)\, dm},
\end{equation}
yielding $f_{*\rightarrow \rm SN}/\avg{m}\simeq1.3\times10^{-2}\;\rm
M_\odot^{-1}$.  If we were to reduce the upper mass limit to 50~\msun\ yields
$f_{*\rightarrow \rm SN}/\avg{m}\simeq1.2\times10^{-2}\;\rm M_\odot^{-1}$, and
an upper mass limit of 20~\msun\ yields $0.9\times10^{-2}\;\rm M_\odot^{-1}$.
The effect of these variations on $\eta$ is unclear, because the upper mass
limit of the IMF also affects our translation of UV and IR light into SFR.
The UV and IR maps are primarily sensitive to high mass stars. Therefore for a
lower upper mass limit, therefore, less UV and IR light is emitted per unit
star formed and if the upper mass limit is actually lower than we have
assumed, then we have underestimated the true SFR. In calculating $\eta$,
these higher SFR and lower $f_{*\rightarrow \rm SN}/\avg{m}$ have opposite
effects, leaving the impact of changing the upper mass limit on $\eta$
unclear. As the upper mass limit decreases, we estimate less intrinsic UV and
IR emission per high-mass star, which we have used to construct the maps. We
neglect the contribution of type Ia SNe whose rate is $\sim1/3$ of the
core-collapse rate for the morphological types of our sample galaxies
\citep{Mannucci2005}.  In these circumstances, the SN rate can be
straightforwardly calculated as a function of SFR, $\eta=\eta(\Sigma_{\rm
  SFR})$, and, depending on the adopted assumptions, the SN rate is
characterized by an uncertainty of a factor $\sim1/3$.  We assume that for
each SN explosion only a fraction $\epsilon_{\rm SN}\le1$ of $10^{51}$~erg --
roughly the energy released by a single SN event \citep{Heiles1987} -- is
converted into turbulence. Then in steady state the kinetic energy of the gas
turbulence $E_k=\eta\times(\epsilon_{\rm SN}\,10^{51}\;{\rm erg})\tau_D$.

\begin{figure*}
\epsscale{1.0}
\plotone{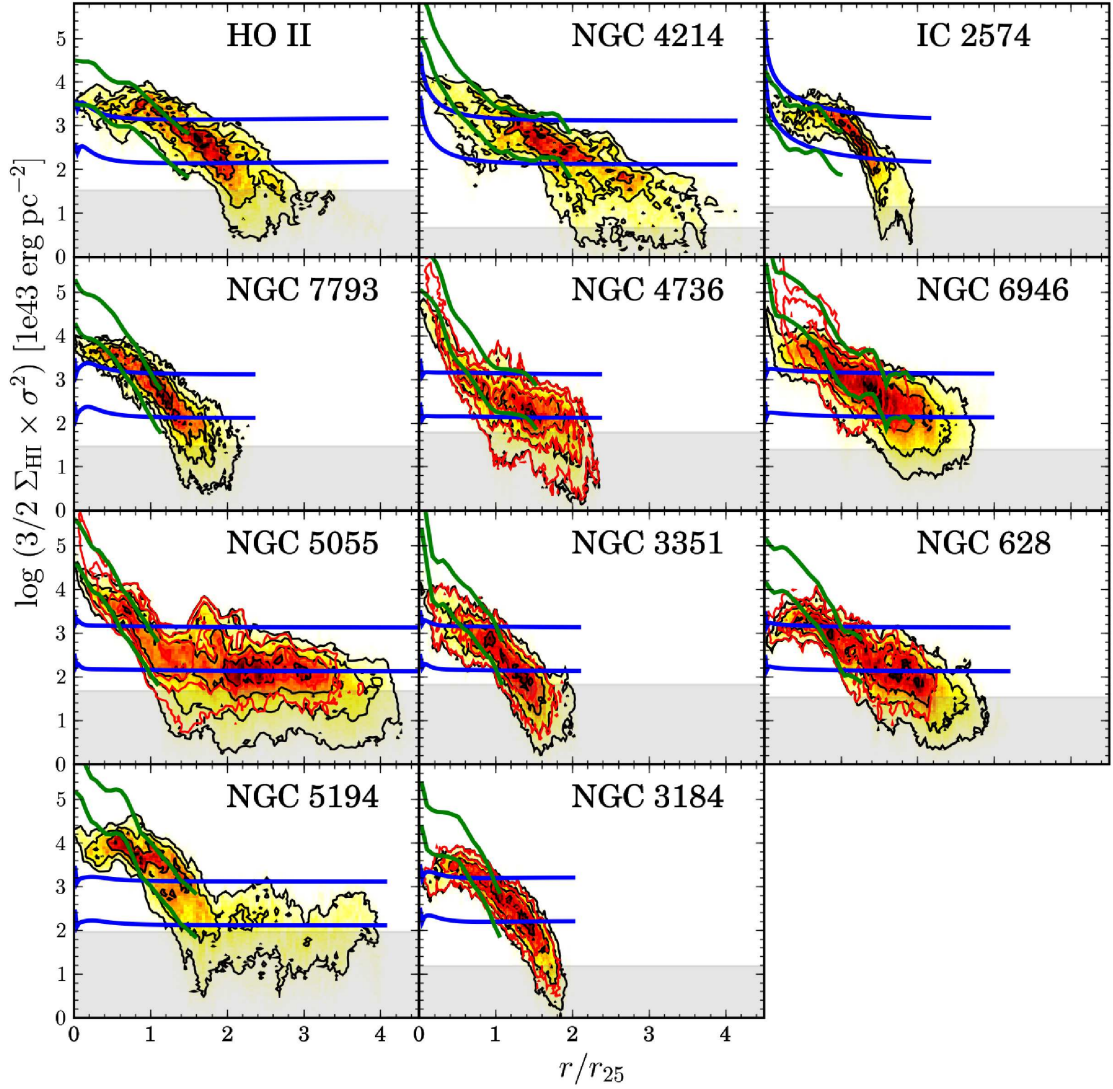}
\caption{ Radial
  distributions of {\sc Hi} kinetic energy per unit area,
  $E_k=3/2$~\shi~\sighi$^2$, for the galaxies of our sample
  (\S~\ref{sec:radialek}).  The number of pixels falling at each position in
  the plot is represented in color scale and contoured in black.  The red
  contours represent the sum of kinetic energies for {\sc Hi} and H$_2$ gas,
  which diverge from the {\sc Hi} contours only near galactic centers.  The
  solid green lines represent the total SN energy released over the
  self-dissipation timescale of turbulence for values of the SN efficiency
  $\epsilon_{\rm SN} = 1$ (above) and~0.1 (below), as discussed in
  \S~\ref{sec:supernova}.  The SN energy estimates plotted here retain only
  those values above the \sfr\ noise threshold, $\Sigma_{\rm SFR}\ge-10\;{\rm
    M_\odot\;yr^{-1}\;pc^{-2}}$ (\S~\ref{sec:limitations}).  The blue
  solid lines indicate MRI energy produced over the corresponding timescale
  for turbulence dissipation for values of the MRI efficiency $\epsilon_{\rm
    MRI}=1$ (above) and~0.1 (below) as discussed in \S~\ref{sec:mri}.  The
  gray horizontal shadow indicates the noise level of $E_k$.
\label{fig:Ek_vs_r}}
\end{figure*}

\begin{figure*}
\epsscale{1.0}
\plotone{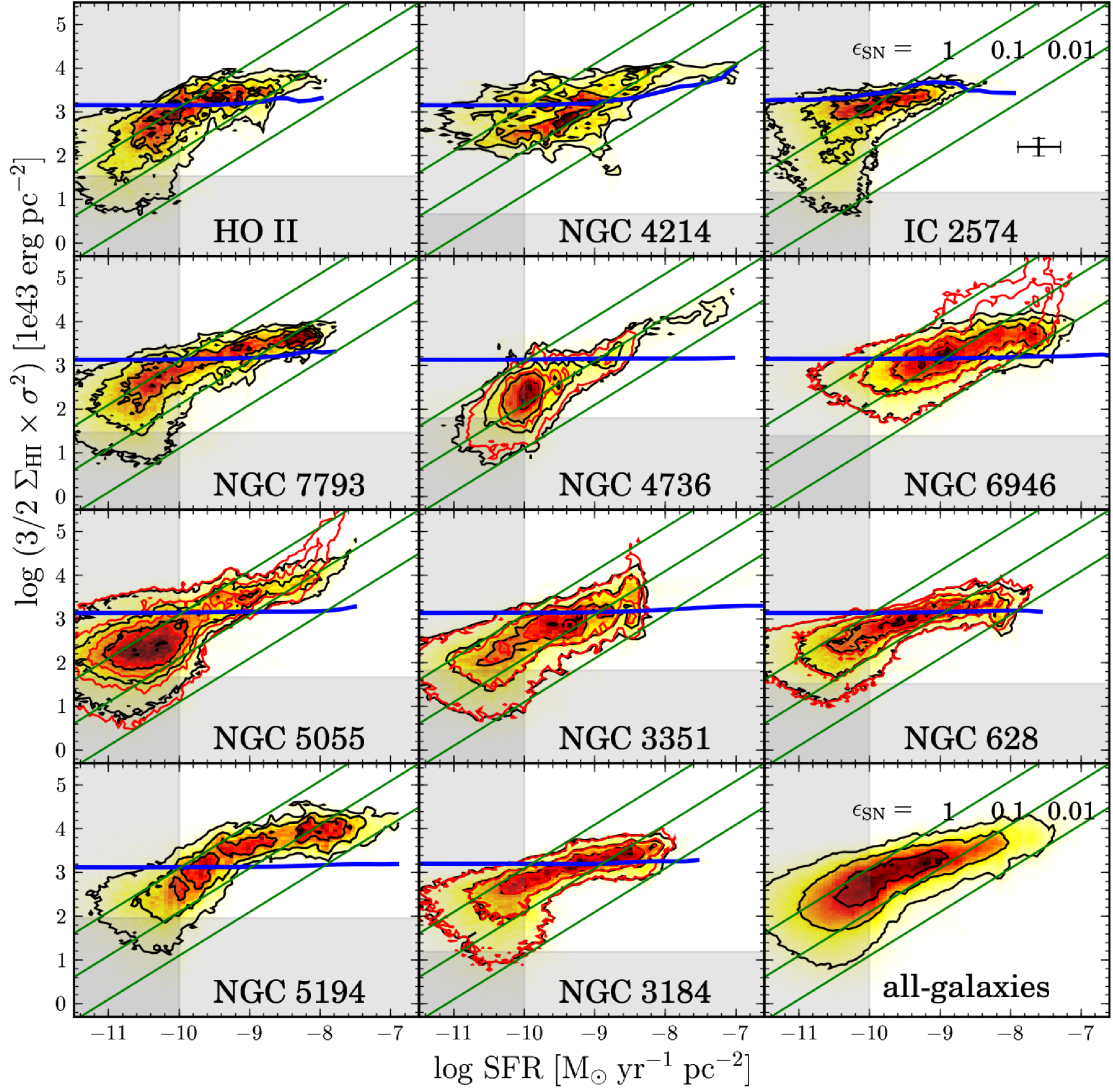}
\caption{Pixel-by-pixel scatter plot of $E_k=3/2$~\shi~\sighi$^2$ vs
  \sfr\ for the galaxies of our sample (\S~\ref{sec:pixel}). The number of
  pixels at each position is represented in color scale and contoured in
  black. The red contours represent the sum of kinetic energies for {\sc Hi}
  and H$_2$ gas.  The solid lines of unity slope represent the SN energy input
  for different SN efficiency values $\epsilon_{\rm SN}=1,\;0.1,\;0.01$ (top
  to bottom; see \S~\ref{sec:supernova}). The blue solid line indicates the
  MRI energy input at maximum efficiency $\epsilon_{\rm MRI}=1$ (see
  \S~\ref{sec:mri}). The error bar of a typical pixel is indicated in the
  upper right panel. The gray vertical and horizontal shadows indicate the
  noise level of \sfr\ and $E_k$, respectively \citep{Leroy2008}. The last
  panel represents the superposition of data points for all the galaxies of
  the sample.\label{fig:Ek_vs_sfr}}
\end{figure*}

In Fig.s~\ref{fig:Ek_vs_r} and~\ref{fig:Ek_vs_sfr}, we compare our estimate
for the total energy from SNe produced within a single turbulent decay time
$\tau_D=9.8$~Myr with $E_k$, the kinetic energy derived from the line width,
as a function of radius and of SFR, respectively, for different values of SN
efficiency.  Fig.~\ref{fig:Ek_vs_r} shows the azimuthally averaged SN energy
decreasing as a function of radius.  A universal $\epsilon_{\rm SN}$ would
produce a linear correlation between $E_k$ and \sfr\ as shown in
Fig.~\ref{fig:Ek_vs_sfr}, where lines of unity slope and constant values of
$\epsilon_{\rm SN}$ represent SN energy input as a function of SFR.
Fig.~\ref{fig:Ek_vs_r} and Fig.~\ref{fig:Ek_vs_sfr}\ show that, at least in
the high star formation regime ($\Sigma_{\rm SFR} >10^{-9}\; {\rm
  M}_{\odot}$~yr$^{-1}$~pc$^{-2}$), the data points typically agree with an
efficiency $0.1\le\epsilon_{\rm SN}\times(10^7\;{\rm yr}/\tau_D)\le1$.

This estimate of the efficiency is consistent with numerical simulations that
estimate $\avg{\epsilon_{\rm SN}}\simeq0.1$ \citep{Thornton1998}.  In the low
SFR regime ($\Sigma_{\rm SFR} < 10^{-9}\; {\rm
  M}_{\odot}$~yr$^{-1}$~pc$^{-2}$), many data points in
Fig.s~\ref{fig:Ek_vs_r} and~\ref{fig:Ek_vs_sfr} lie close to the line of
maximum efficiency $\epsilon_{\rm SN}=1$.  At $\Sigma_{\rm SFR} <
10^{-10}\;{\rm M}_{\odot}$~yr$^{-1}$~pc$^{-2}$, in particular at large
galactocentric radii, the estimated \sfr\ rates fall to within the noise (see
\S~\ref{sec:limitations}), whilst the measured $E_k$ is typically well
constrained.  Nevertheless, $\epsilon_{\rm SN} = 1$ remains problematic, since
it would imply that all the kinetic energy of SN remnants would be deposited
as kinetic energy of the gas.  In fact \citet{TenorioTagle1991}\ argue that
the expected SN efficiency $\epsilon_{\rm SN}$ should be at most $\sim0.5$.
Values graeter than 0.5, therefore, either imply that other sources inject
energy into the ISM (see following sections), or that the dissipation
timescales are shorter than we assume. In the next sections, we examine
alternative mechanisms to explain the line width in these regions.

\subsection{Thermal Broadening}\label{sec:thermal}

The temperature of the warm phase of the \hi\ is maintained by UV heating.
However, neutral gas never reaches temperatures high enough to explain line
widths as high as 10~\kms, which are instead attributed to supersonic
turbulent motions \citep{Wolfire2003, Lacour2005}.  As
\sighi~$\gtrsim10$~\kms\ for $r\lesssim r_{25}$ for all our sample galaxies,
we argue that thermal effects can be neglected within the star forming radius.
In regions of active star formation, stellar winds and ionizing radiation from
stars appear less effective together than the SNe from the same stellar
population at driving turbulence \citep{maclow2004}. Thus, SN driven
turbulence looks likely to dominate there.

In the outer regions of \hi\ disks, on the other hand, the velocity dispersion
approaches its thermal value.  There, the warm neutral atomic phase
temperature is typically $\sim5000$~K, as measured, e.g., in the solar
neighborhood \citep{Heiles2003, Redfield2004}, where the SFR is a few
$\times10^{-9}\; \rm M_\odot\;yr^{-1}\;pc^{-2}$.  This temperature gives a
thermal width of \sighi~$\sim6$~\kms. Typical temperatures of the warm medium
can even be as high as $\sim8500$~K \citep[corresponding to
$\sim8$~\kms;][]{Wolfire1995}.  If thermal broadening is effective, the
observed \sighi~$\sim6$~\kms\ outside the star forming radius does not
necessarily involve turbulence.  However, such temperature levels, especially
at $r\gtrsim2\times r_{25}$, where the SFR is below $10^{-10}\; \rm
M_\odot\;yr^{-1}\;pc^{-2}$ (cf.  Fig.~\ref{fig:m2_vs_r_fit}), still require a
continuous UV background source warming up the \hi\ disks of galaxies. At
$r>r_{25}$, the source of such UV radiation is presumably not local; it could
be extragalactic.

The local thermal pressure estimate and the actual velocity dispersion in the
outer parts of \hi\ disks are consistent with the existence of a warm phase
for the gas.  For example, \citep{Leroy2008} estimate a local pressure
$P/k\sim300$~K~cm$^{-3}$ at $r\simeq2\times r_{25}$ for the galaxy NGC~4214,
assuming hydrostatic equilibrium \citep[cf.][]{Elmegreen1989}.  At a radius
$2\times r_{25}$, the \hi\ velocity dispersion in NGC~4214 is $\sim7$~\kms,
corresponding to a temperature of $\sim5900$~K, which, solving $P\propto
n\,\sigma^2$, yields a density $n\simeq0.05$~cm$^{-3}$. This is consistent
with the gas in a warm phase according to Figure~7 in \citet{Wolfire2003},
where a pressure of $\sim300$~K~cm$^{-3}$ corresponds to
$n\sim0.04$~cm$^{-3}$, at least for the outer parts ($r\sim 15-18$~kpc) of the
Milky Way, which likely have a similar UV background.  For pressure values
lower than $300$~K~cm$^{-3}$, all the gas is warm.

\subsection{MRI Energy}\label{sec:mri}

If external UV heating is insufficient to maintain a warm phase, then a
non-stellar energy source is needed to explain the observed turbulence
(\S~\ref{sec:radialek}). MRI is a plausible candidate. It develops in
differentially rotating disks with angular velocity decreasing outwards, as in
all but the smallest galactic disks, as long as some weak magnetic field
threads the disk.  It has been argued to sustain both the interstellar gas
turbulence and the galactic magnetic field to a few $\mu$G \citep{Piontek2005,
  Piontek2007}. MRI requires a minimum magnetic field as low as $10^{-25}$~G
to originate, which is much lower than the seed galactic magnetic fields
\citep{Kitchatinov2004}.

Following the same line of reasoning as \S~\ref{sec:supernova}, if we assume
that a fraction $\epsilon_{\rm MRI}\le1$ of the MRI energy is transformed into
kinetic energy, then in steady state the observed kinetic energy must equal
the MRI energy input within the turbulence decay time: $E_k=\epsilon_{\rm
  MRI}\,\dot{E}_{\rm MRI}\;
\tau_{\rm MRI}$. Theoretical calculations \citep{Sellwood1999, maclow2004,
  maclow2008} estimate the production energy rate of MRI to be
\begin{equation}
\begin{array}{cl}
\dot{E}_{\rm MRI}= &
3.7\times10^{-8}\;{\rm erg\;cm^{-2}\;s^{-1}}\times \\
 & \;\left(\frac{h_z}{100\;{\rm pc}}\right)\;
 \left(\frac{B}{6\;\mu{\rm G}}\right)^2\;\frac{\Omega}{(220\;{\rm Myr})^{-1}},\\
\end{array}
\end{equation}
where $\Omega\equiv v/r$ is the angular velocity, $h_z$ is the vertical
thickness of the \hi\ disk, and $B$ is the magnetic field.  In our analysis,
we assume a constant thickness $h_z=100$~pc for all galaxies. This is
appropriate for the inner Milky Way out to the solar circle
\citep{Wolfire2003}, although in the outer regions of disks and in dwarf
galaxies $h_z$ may be higher \citep[up to $\sim300$~pc;][]{Walterbrinks1999}.
We also assume a constant magnetic field $B=6\;\mu{\rm G}$ as a typical
galactic magnetic field \citep{Beck1996, Heiles2005}. Taking the turbulent
decay timescale again to be $\tau_{\rm MRI}
\simeq9.8\;(\lambda_{100}/\sigma_{10})\;\rm Myr$ we can estimate the driving
scale $\lambda=\lambda_c$.  The critical wavelength for the fastest MRI growth
\citep{Balbus1998}
\begin{equation}
\lambda_c=2\pi\:v_A\; 
\left[ -\frac{3+\alpha}{4}\;\frac{d\Omega^2}{d\ln r}\right]^{-1/2},
\end{equation}
where $\alpha\equiv d\ln v/d\ln r$, and the Alfv\'en velocity $v_A^2 = B^2 / 4
\pi \rho$.  For a typical galactic magnetic field of $B=6\;\mu{\rm G}$ and a
density of $2\times10^{-24}$~g~cm$^{-3}$, $\lambda_c\sim10^2$~pc.

In Fig.~\ref{fig:Ek_vs_r}, we compare the energy produced by MRI within a
decay time of turbulence, $\epsilon_{\rm MRI}\,\dot{E}_{\rm MRI}\,\tau_{\rm
  MRI}$, with the observed kinetic energy $E_k$ as a function of radius for
different values of the MRI efficiency. Note that although both $\dot{E}_{\rm
  MRI}$ and $\tau_{\rm MRI}$ are functions of radius, the MRI energy input
only decreases with radius slowly, not exponentially as $E_k$. In
Fig.~\ref{fig:Ek_vs_r}, we show a comparison between the SN energy
(\S~\ref{sec:supernova}) and the MRI energy plotted as a function of radius,
and indicated with green and blue solid lines, respectively. We also plot in
the pixel-by-pixel $E_k$ vs.\ \sfr\ plot of Figure~\ref{fig:Ek_vs_sfr}\ the
average value of the MRI for pixels with each \sfr, for $\epsilon_{\rm
  MRI}=1$.
    
While SNe can be identified as the dominant source of energy at $\Sigma_{\rm
  SFR}>-9\;{\rm M_\odot\;yr^{-1}\;pc^{-2}}$, the MRI contribution becomes
important at $\Sigma_{\rm SFR}<-9\;{\rm M_\odot\;yr^{-1}\;pc^{-2}}$ and
dominant at large galactocentric radii. On the other hand, the kinetic energy
of turbulence within $r_{25}$ is much higher than the predicted MRI energy,
indicating that MRI can not account for the observed gas turbulence in regions
of active star formation.

The observed $E_k$ is higher than the SN energy for the dwarf galaxies
Holmberg~II and, more significantly, IC~2574. In these two cases, the gas
turbulence cannot be explained by SNe \citep[cf.][]{Stanimirovic2001, Dib2005,
  Pasquali2008}; still, the MRI could account for the observed regime of
turbulence.  Our analysis suggests that MRI could dominate regions of low SFR,
and rules out MRI as an effective turbulence driving mechanism at high SFR.

\subsection{Robustness  of the Approach}\label{sec:errors}

Three major sources of uncertainty enter our analysis.  The first and largest
of these are the empirical conversions from UV and IR emission to \sfr, which
may introduce up to a 40\% uncertainty (\S~\ref{sec:limitations}). The
conversion from \sfr\ to SN rate relying on a universal IMF introduces an
additional 30\% uncertainty (\S~\ref{sec:supernova}).  A second uncertainty
enters from the reliance on numerical simulation to evaluate the driving scale
of SN-driven turbulence and thus the dissipation timescale $\tau_{\rm SN}$.
The simulations give a 30\% error for their estimate of the driving scale
\citep{Joung2006, Avillez2007}.  The dissipation timescale also depends on the
mean velocity dispersion, which is, of course, not constant as assumed in our
estimate. A third uncertainty enters in our estimates of MRI energy, where we
assume a constant magnetic field and vertical thickness of the disk for all
the galaxies of the sample, which may not be the case. The vertical thickness
of the gaseous component in galaxies increases outwards and may double the
value assumed in our analysis of $h_z=100$~pc. The typical strength of
magnetic fields observed in the Milky Way and other galaxies declines slowly
as a function of galactocentric distance, although the variations of the
magnetic field can be much larger azimuthally than radially
\citep{Fletcher2004, Han2006, Beck2007}.  These three sources of errors could
in principle explain those data points lying near $\epsilon_{\rm SN}\gtrsim1$
and $\epsilon_{\rm MRI}\gtrsim1$.

Aside from the potential sources of uncertainties, if we were to interpret our
empirical finding by taking into account only SNe explosions and MRI, still a
minor part of the observed turbulence would require $\epsilon_{\rm SN}$ and
$\epsilon_{\rm MRI}$ efficiencies uncomfortably high, as high as 100\%.
Therefore, we do not exclude that other mechanisms could be efficiently
driving some turbulence in the gas. Potentially, within the star forming
radius ($\lesssim r_{25}$) stellar winds could be the most effective, while in
the outermost regions of \hi\ disks, i.e. $r\gtrsim2\times r_{25}$, where the
SNe and the star formation effects are not likely to produce feedback, a floor
level of \hi\ velocity dispersion of $\sim6$~\kms\ could be attributed to
thermal broadening.  Observation of the \hi\ in absorption at large
galactocentric radii might ultimately help in determining whether the gas is
cold and turbulent or warm \citep[see][]{Dickey1993}.

\subsection{Is \shi\  Controlling the $E_k$ vs \sfr\ Relation?}\label{sec:controlling}

\begin{figure*}
\epsscale{1.0}
\plotone{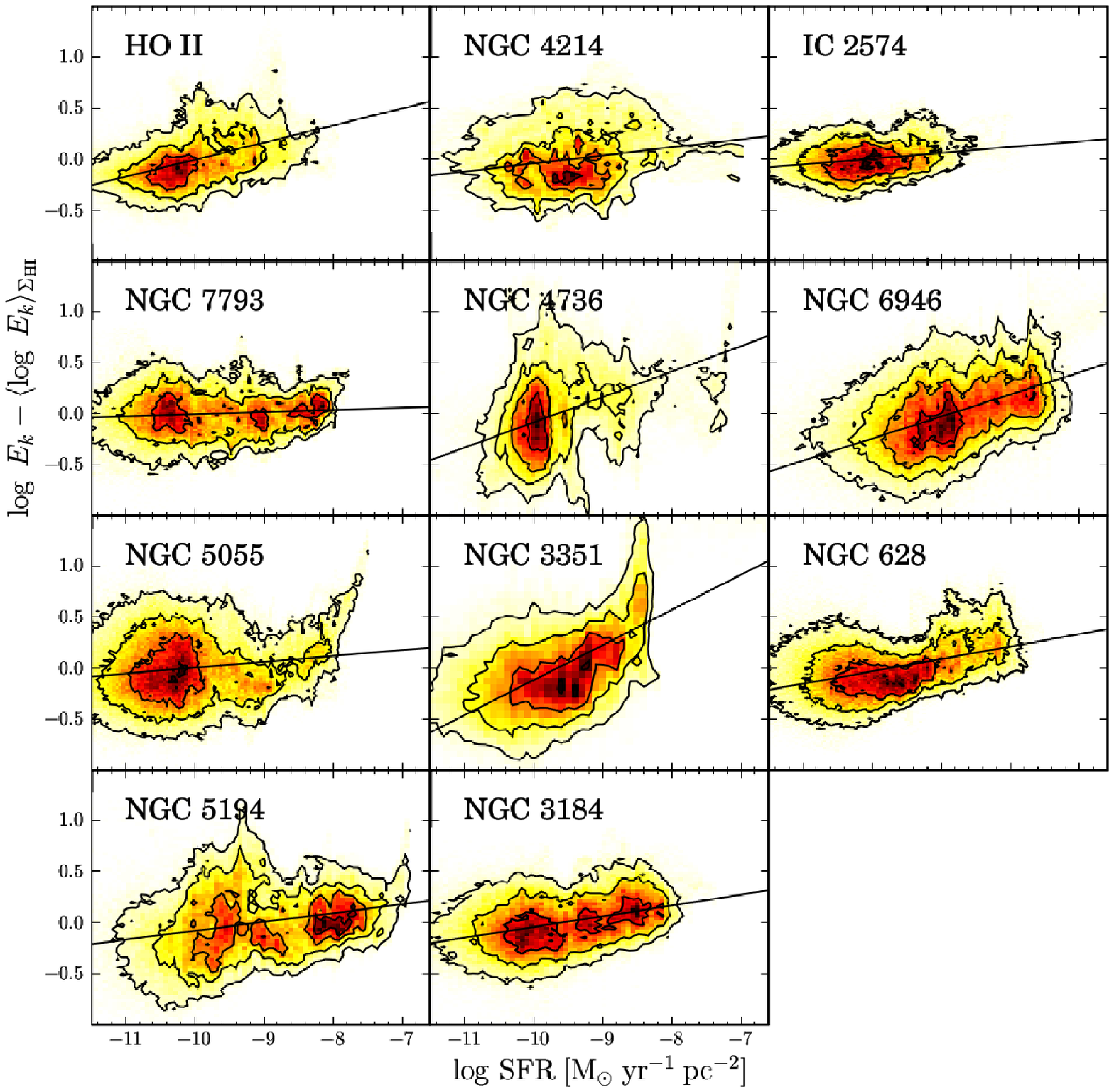}
     \caption{Pixel-by-pixel scatter plot of residuals $E_k
       -\avg{E_k}_{\Sigma \rm HI}$ vs \sfr\ after subtracting the average of
       $E_k$ at constant \shi. The solid lines represent the linear fit to
       data. This procedure is equivalent to looking at the correlation
       between $E_k$ and \sfr\ at constant \shi. The residuals are positively
       correlated with \sfr, indicating that \shi\ does not determine the
       observed $E_k$ vs \sfr\ correlation alone (\S~\ref{sec:controlling}).
       Beside this graphical illustration, we perform the analytical test of
       the actual correlation between $E_k$ and \sfr\ while controlling for
       \shi\ through calculating the partial correlation coefficent, as
       described in the text (see also Table~\ref{tab:corrs} for a list of the
       partial correlation coefficients).  \label{fig:Ek_res_vs_sfr}}
\end{figure*}

The gas kinetic energy, $E_k$, is the correct physical quantity to study the
energy balance in the ISM. However, gas surface density and SFR are well known
to correlate \citep[e.g.][]{Kennicutt1998, Bigiel2008}, so one may wonder
whether this drives our observed correlation between $E_k$ and \sfr.  In other
words, the observed covariation between $E_k$ and \sfr\ might result from the
two facts: at higher gas mass density, galaxies form stars at higher rate, and
higher gas mass bears higher kinetic energy. In order to verify that the
correlation between $E_k$ and \sfr\ is not controlled by \shi, we remove the
effect of \shi\ on the $E_k$ vs \sfr\ relation by calculating the partial
correlation coefficient:
\begin{equation}\label{eq:partcorr}
\rho_{12.3} = \frac { \rho_{12} - \rho_{13} \rho_{23} }
 {\sqrt{ (1-\rho^2_{13}) (1-\rho^2_{23}) }},
\end{equation}
where $\rho_{12.3}$ is the partial correlation between $x_1\equiv\log
\Sigma_{\rm SFR}$ and $x_2\equiv\log E_k$ while controlling for $x_3\equiv\log
\Sigma_{\rm HI}$, and $\rho_{ij}$ is the Pearson's correlation coefficient
between two data sets $x_i$ and $x_j$. Here, considering the quantities in
logarithmic scale is convenient as they are correlated as power laws.
Retaining only data points above a fiducial value for the \hi\ mass density,
i.e.  $\Sigma_{\rm HI}\ge 3$~\msun\ \citep[cf.][]{Bigiel2008}, we obtain the
values listed in Table~\ref{tab:corrs} for our sample galaxies.  If the
correlation between $E_k$ and \sfr\ were completely controlled by \shi\ we
would expect $\rho_{12.3} = 0$.  Yet, we find that $0.2\le\rho_{12.3}\le0.6$
for our sample galaxies, indicating that the correlation between $E_k$ and
\sfr\ is real.  Equivalently, we show that the correlation $E_k$ vs \sfr\ at
constant \shi\ holds a positive slope in Fig.\ref{fig:Ek_res_vs_sfr}, in which
we remove the contribution from \shi\ to $E_k$ by subtracting the average
$\avg{E_k}_{\Sigma \rm HI}$ within bins of \shi. The residuals $E_k
-\avg{E_k}_{\Sigma \rm HI}$ vs \sfr\ exhibit a positive correlation,
indicating that it is not \shi\ alone that determines the observed $E_k$ vs
\sfr\ correlation.  Positive slopes in Fig.~\ref{fig:Ek_res_vs_sfr} and
positive partial correlation coefficients imply that there is a real, physical
relationship between $E_k$ and SFR even at fixed \shi. The relatively weak
slopes in Fig.~\ref{fig:Ek_res_vs_sfr} indicate that higher gas mass density
correlates indeed with higher kinetic energy, simply because it generates more
star formation.  Fig.~\ref{fig:Ek_res_vs_sfr}\ and the partial correlation
coefficients allow us to detect a relationship between $E_k$ and SFR that is
independent of \shi. Comparing Fig.~\ref{fig:Ek_vs_sfr}\ and
Fig.~\ref{fig:Ek_res_vs_sfr}, however, it is clear that most of the
correlation between $E_k$ and SFR in Fig.~\ref{fig:Ek_res_vs_sfr} closely
involves \shi\ (the distributions in Fig.~\ref{fig:Ek_res_vs_sfr}\ are very
flat compared to those in Fig.~\ref{fig:Ek_vs_sfr}). The basic effect seems to
be that higher \shi\ results in higher SFR, which creates more $E_k$.  Also,
note that if the turbulence, as traced by $E_k$, were effectively suppressing
star formation, we would have observed a negative correlation here.

\subsection{Does Turbulence Drive Stochastic Star Formation?}

The data analysis suggests that the SFR drives the \hi\ turbulence through SN
feedback. However, the observed correlation could also be interpreted in the
opposite logical direction, i.e., that the turbulence is driving star
formation. In fact, as it has been argued \citep{maclow2004}, turbulence in
the ISM has a dual role: (1) to quench star formation by providing pressure
support to the ISM and preventing collapse, and (2) to promote star formation
by generating stochastic super-critical density enhancements. If the
turbulence were to drive substantial stochastic star formation, we would
indeed expect a positive correlation between \sighi\ and \sfr. However,
Fig.~\ref{fig:m2_vs_r_fit}\ shows that \sighi\ and \sfr\ occupy quite
different dynamic ranges.  While the \sfr\ ranges over several orders of
magnitude, \sighi\ ranges from $\sim20$ to $\sim5$~\kms\ and is characterized
by large azimuthal variations.  Although \sfr\ positively correlates with
\sighi\ on galactic scales, the large azimuthal variations imply that \sfr\ is
not well defined for any given value of \sighi.  This does not preclude
turbulent induction of star formation in individual regions, but does suggest
that this process does not dominate over large scales.  The physical
explanation might be that supercritical density fluctuations are often
dispersed on timescales shorter than the free-fall time, arresting the
collapse \citep{Klessen2000, Elmegreen2002, Joung2006}.

\section{Effects of Spiral Arm Kinematics and Tidal Interactions}

In the following we discuss other possible mechanisms to produce ISM
turbulence, such as  spiral arm
kinematics, tidal interactions, and streamers.

Table~\ref{tab:objs}, which lists the morphological types of the galaxies of
our sample, shows that there is no evident trend of the mean velocity
dispersion, $\avg{\sigma}$, among individual galaxies or morphological type.
Spiral galaxies with strong spiral pattern, e.g. NGC~628 and NGC~3184, have
similar values of typical \hi\ velocity dispersion as galaxies with no clear
spiral structure, e.g. Holmberg~II and IC~2574. Since the spiral arm strength
should vary within the sample, we argue that the spiral arm kinematics in our
sample galaxies are not an important effect in driving turbulence into the
ISM.

Although the galaxies Holmberg~II and IC~2574 belong to the M81 
group, they do not show signatures of tidal distortion. In our sample, only NGC~5194 is an interacting galaxy. The \hi\ velocity dispersion
in NGC~5194 is significantly higher than the average for the galaxies in the
sample (see Fig.~\ref{fig:s25_avg}).  On the basis of our results, we
speculate that  the tidal interaction with the companion NGC~5195
 enhanced the SFR in the disk of NGC~5194, which has
consequently driven the velocity dispersion in the \hi\ gas to higher values.

Extended streamers characterize the galaxies NCG~4736 and NGC~5055.  Their
radial profiles of the \hi\ velocity dispersion exhibit a local increase
outside the radius of active star formation, i.e., at $r\sim4'$ ($\sim
r_{25}$) for NCG~4736 and at $r\sim9'$ ($\sim1.5\, r_{25}$) for NGC~5055.
However, these local peaks in \sighi($r$) and the streamers have different
galactocentric locations, corresponding to $r\sim8'$ and $r>11'$ for NCG~4736
and NGC~5055, respectively. Therefore, we argue that the presence of extended
streamers is not likely to be connected to higher \hi\ velocity dispersion.

\section{CONCLUSIONS}

Combining high quality maps of \hi\ column density and line width provided by
THINGS for a sample of dwarf and spiral galaxies, we obtain the following
results.
\begin{enumerate}
\item The \hi\ velocity dispersion, \sighi, declines uniformly as a function
  of galactocentric distance in all analyzed galaxies.
  
\item At $r_{25}$, the edge of the star-forming region, the \hi\ velocity
  dispersion $\sigma(r_{25})\simeq10\pm2$~\kms, which is consistent with the
  mass-weighted median \hi\ velocity dispersion $\avg{\sigma}$. These findings
  are independent of the dynamical mass of the galaxy and of their
  morphological type.
  
\item Within the radius of active star formation ($r\le r_{25}$), the
  estimated SN rate and the corresponding energy input rate are sufficient to
  account for the bulk of observed kinetic energy of turbulence.  For those
  galaxies of the sample with considerable H$_2$ gas, the SNe can well account
  for the combined \hi\ and H$_2$ gas turbulence.  In this region, the
  observed instantaneous kinetic energy of the \hi\ gas is consistent with the
  balance between the energy input from the total number of SNe calculated
  from the observed SFR and the turbulent dissipation predicted by numerical
  models.  The proportionality between gas $E_k$ and SN energy input rate
  derived from the SFR provides direct evidence that \hi\ turbulence comes
  from SNe in regions of active SFR. The resulting SN feedback efficiencies
  are typically $\epsilon_{\rm SN}\times(10^7\;{\rm yr}/\tau_D)\simeq0.1$ at
  SFR levels $\Sigma_{\rm SFR} >10^{-9}$~M$_{\odot}$~yr$^{-1}$~pc$^{-2}$, with
  the dissipation timescale of turbulence $\tau_D\simeq10^7$~yr.
  
\item Within the star forming disk ($r\le r_{25}$), neither thermal broadening
  nor MRI can produce the observed \hi\ velocity dispersion.  At low SFR,
  $\Sigma_{\rm SFR} <10^{-9}$~M$_{\odot}$~yr$^{-1}$~pc$^{-2}$, corresponding
  to large radial distances ($r>r_{25}$), an additional mechanism driving the
  \hi\ velocity dispersion is required to avoid SN efficiencies $\epsilon_{\rm
    SN} > 1$.
  
\item The thermal broadening of the spectral lines, associated to a
  temperature of $\sim5000$~K, may be able to explain the observed
  $\sigma_{\rm HI}\sim 6$~\kms\ in the outermost regions of \hi\ disks in our
  sample galaxies, if the required UV radiation to maintain these temperatures
  is present.  The energy input from MRI can account for the kinetic energy
  observed in regions of low SFR, $\Sigma_{\rm SFR}
  <10^{-9}$~M$_{\odot}$~yr$^{-1}$~pc$^{-2}$, at large galactocentric
  distances.
  
\item We can not unambiguously separate the temperature of the warm and
  non-turbulent neutral medium from the effect of MRI stirring the ISM.  Both
  mechanisms are equivalently plausible drivers of the \hi\ velocity
  dispersion observed in the outer parts ($r>r_{25}$) of galaxy disks. We
  suggest that testing the \hi\ line profiles of the gas against a bright
  background source could ultimately clarify whether the gas in regions of
  weak star formation is uniformly warm, or contains a cold, turbulent phase,
  presumably stirred by MRI.  If the gas is actually turbulent, the gas
  kinetic energy for both high and low star forming regions is consistent in
  all cases with realistic values of $\epsilon_{\rm SN}$ and $\epsilon_{\rm
    MRI}$ efficiencies, suggesting that the feedback provided by both SN
  explosions and MRI is sufficient to drive the bulk of the observed \hi\ 
  turbulence.

\end{enumerate}

\subsubsection*{Acknowledgments}

We are grateful to the anonymous referee for the valuable and
interesting comments that improved the quality of the paper.
We acknowledge helpful discussions with E. Bell, S. Dib, N.
Dziourkevitch, R. Klessen, and A. Pasquali.  M-MML thanks the
Max-Planck-Gesellschaft for support during his visit to the MPIA.

\renewcommand{\baselinestretch}{1}

 \clearpage

\clearpage

\begin{table*}
 \begin{center}
\begin{tabular}{lccccccccc}
\tableline\tableline
Obj. name & D     & $i$   & $r_{25}$ & $\delta v$ & $\langle\sigma\rangle$ & $\sigma_{25}$ & $\partial\sigma/\partial r$  & morph. & morph. \\
          & (Mpc) & (\dg) &  ($'$)   &  (\kms)    & (\kms)                 &  (\kms)       & (\kms\ $r_{25}^{-1}$)   & code & type  \\
          & (1)   & (2)   & (3)      & (4)        &  (5)                   &  (6)          &   (7)  & (8)  & (9) \\
\tableline \tableline                                                                                      
\multicolumn{8}{c}{Dwarf galaxies} \\
\tableline
HO II     & 3.39  & 41    &   3.3    & 2.6        & 9.4                    &  $9.9\pm0.5$  & $-2.5\pm0.5$ & 10 & Irr \\
IC 2574   & 4.02  & 53.4  &   6.44   & 2.6        & 9.7                    &  $10.8\pm0.7$ & $-6.1\pm0.7$ & 9 & Sm  \\
NGC 4214  & 2.94  & 43.7  &   3.38   & 1.3        & 8.9                    &  $9.2\pm0.3$  & $-2.1\pm0.3$ & 10 & Irr  \\
\tableline\tableline                                                                                        
\multicolumn{8}{c}{Normal spiral galaxies} \\                                                                 
\tableline                                                                                                 
NGC 628   & 7.3   & 7     &   4.77   & 2.6        & 8.0                    &  $7.7\pm0.4$  & $-1.8\pm0.4$ & 5 & S?  \\
NGC 3184  & 11.1  & 16    &   3.62   & 2.6        & 10.4                   &  $9.7\pm0.9$  & $-5.2\pm0.9$ & 6 & Sc \\
NGC 3351  & 9.33  & 41    &   3.54   & 5.2        & 10.6                   &  $9.6\pm0.9$  & $-2.6\pm0.9$ & 3 & Sb \\
                                                                                                                 
NGC 4736  & 4.66  & 41.4  &   3.88   & 5.2        & 12.0                   &  $12.1\pm1.4$ & $-4.7\pm1.4$ & 2 & Sab \\
NGC 5055  & 7.82  & 59    &   6.01   & 5.2        & 12.6                   &  $11.4\pm0.3$ & $-1.2\pm0.3$ & 4 & Sbc \\
NGC 5194  & 7.77  & 42    &   4.89   & 5.2        & 17.7                   &  $17.0\pm1.4$ & $-3.3\pm1.4$ & 4 & Sbc \\
NGC 6946  & 5.5   & 32.6  &   5.35   & 2.6        & 10.1                   &  $8.6\pm0.9$  & $-1.0\pm0.9$ & 6 & Sc \\
NGC 7793  & 3.82  & 50    &   5.0    & 2.6        & 11.4                   &  $9.7\pm1.1$  & $-2.6\pm1.1$ & 7 & Scd \\

\tableline\tableline
\end{tabular}\caption{THINGS target galaxies. (1) adopted distance
  \citep{walter2008}; (2) inclination \citep{deblok2008}; (3) semi-major axis
  of the 25~mag~arcsec$^{-2}$ isophote in the $B$ band obtained from the LEDA
  database (URL: http://leda.univ-lyon1.fr/); (4) velocity resolution of the
  {\sc Hi} data cubes; (5) {\sc Hi} mass-weighted median of the {\sc Hi}
  velocity dispersion; (6) {\sc Hi} velocity dispersion $\sigma(r_{25})$ at
  $r_{25}$ (\S~\ref{sec:radial}; figure caption Fig.~\ref{fig:s25_avg}); (7)
  slopes of radial \sighi\ profiles in units of $r_{25}$
  (\S~\ref{sec:radial}); (8) morphological code for the revised de Vaucouleurs
  type (LEDA); (9) morphological Hubble type (LEDA).\label{tab:objs}}
\end{center}
\end{table*}

\begin{table*}
\begin{center}
\begin{tabular}{lc}
\tableline\tableline
object & $\rho_{12.3}$  \\
\tableline
   HO II   &  0.54 \\
NGC 4214   &  0.32 \\
 IC 2574   &  0.19 \\
NGC 7793   &  0.17 \\
NGC 4736   &  0.58 \\
NGC 6946   &  0.52 \\
NGC 5055   &  0.26 \\
NGC 3351   &  0.62 \\
 NGC 628   &  0.52 \\
NGC 5194   &  0.44 \\
NGC 3184   &  0.49 \\
\tableline
\end{tabular}
\caption{Partial correlation coefficients calculated in order to test  the
  actual correlation between $E_k$ and \sfr\ while controlling for \shi\ 
  (\S~\ref{sec:controlling}).\label{tab:corrs}}
\end{center}
\end{table*}

\end{document}